\documentclass[preprints,communication,accept,pdftex,moreauthors]{Definitions/mdpi} 
\firstpage{1} 
\pubvolume{1}
\issuenum{1}
\articlenumber{0}
\pubyear{2023}
\copyrightyear{2023}
\externaleditor{Academic Editors: Prasad S. Thenkabail and Giancarlo Bellucci}
\datereceived{29 May 2023} 
\daterevised{22 August 2023} % Only for the journal Acoustics
\dateaccepted{29 August 2023} 
\datepublished{} 
\hreflink{https://doi.org/} % If needed use \linebreak
\Title{Probing %MDPI: We changed article type to ``communication'' according toour sutem. Please confirm.
Dust and Water in Martian Atmosphere with Far-Infrared Frequency Spacecraft Occultation}

\TitleCitation{Probing Dust and Water in Martian Atmosphere with Far-Infrared Frequency Spacecraft Occultation}

\Author{Ananyo Bhattacharya%MDPI: 1. Please carefully check the accuracy of names and affiliations; 2. We added special symbols († ‡ for Ananyo Bhattacharya; ‡ for Cheng Li). Please confirm.
 $^{1,}$*$^{,\dagger,\ddagger}$, Cheng Li $^{1,\ddagger}$, Nilton O. Renno %MDPI: The name of this author is different from the one submitted online at susy.mdpi.com (Nilton O Renno). Please confirm which one is correct. We suggest keeping the current one with abbr. dot.
 $^{1}$, Sushil K. Atreya $^{1}$ and  David Sweeney $^{1,2}$}

\AuthorNames{Ananyo Bhattacharya, Cheng Li, Nilton O. Renno, Sushil K. Atreya, David Sweeney}

\AuthorCitation{Bhattacharya, A.; Li, C.; Renno, N.O.; Atreya, S.K.; Sweeney, D.}
\address{%
$^{1}$ \quad University %MDPI: For universities, the department/school/faculty/campus is required. Please try to provide this information.
 of Michigan, Ann Arbor, MI 48109, USA%MDPI: We added city, state, zip code and country. Please confirm.
; chengcli@umich.edu (C.L.); nrenno@umich.edu (N.O.R.); atreya@umich.edu (S.K.A.); sweendav@umich.edu (D.S.) %MDPI: We added the email addresses here according to those submitted online at susy.mdpi.com. Please confirm.
\\
$^{2}$ \quad NASA Jet Propulsion Laboratory, California Institute of Technology, Pasadena, CA 91125, USA %MDPI: Please check if it is correct.

}

\corres{Correspondence: ananyo@umich.edu}

\firstnote{Current address: Climate and Space Research Building, 2455 Hayward St.
, Ann Arbor, MI 48109, USA.
} 
\secondnote{These authors contributed equally to this work.}
\abstract{Airborne dust plays an active role in determining the thermal structure and chemical composition of the present-day atmosphere of Mars and possibly the planet's climate evolution over time through radiative--convective and cloud microphysics processes. Thus, accurate measurements of the distribution and variability of dust are required. Observations from the Mars Global Surveyor/Thermal Emission Spectrometer Mars Mars Reconnaissance Orbiter/Mars Climate Sounder and Mars Express/Fourier Transform Spectrometer and the Curiosity Rover have limited capability to measure dust. We show that spacecraft occultation of the Martian atmosphere at far-infrared frequencies between 1 and 10 THz can provide the needed global and temporal data on atmospheric dust by providing co-located measurements of temperature and dust opacity from the top of the atmosphere all the way down to the surface. In addition, spacecraft occultation by a small-satellite constellation could provide global measurements of the development of dust storms.}


\begin{document}

%%%%%%%%%%%%%%%%%%%%%%%%%%%%%%%%%%%%%%%%%%

\section{Introduction}

Dust activity in the Martian atmosphere is an important component of the radiative--dynamic coupling processes~\cite{guzewich2013impact,madeleine2011revisiting,medvedev2011influence}. Measurements from landers and orbiters indicate that dust activity occurs at many length scales~\cite{martin1993analysis,clancy2010extension,guzewich2013high,guzewich2013impact,haberle2017atmosphere}. Dust storms influence regional and global climate and weather processes, including seasonal and annual variations of radiative and dynamical properties~\cite{guzewich2013impact}. Dust activity occurs on length and time scales ranging from those of small transient dust devils to seasonal global dust storms. Contact electrification processes in dust devils and dust storms might also trigger electrochemical reactions that impact the composition of the atmosphere~\cite{atreya2006oxidant}. Furthermore, dust activity could modulate the cloud formation of water~\cite{shaposhnikov2018modeling} and CO$_{2}$, as well as the destruction of atmospheric ozone~\cite{daerden2022planet}. Therefore, dust activity likely couples radiative--dynamical feedback, cloud microphysics~\cite{montmessin2002new}, and chemical processes on Mars~\cite{lefevre2008heterogeneous}. Moreover, knowledge of dust activity and the thermal structure of the atmosphere is crucial to understanding the evolution of the atmosphere and to allow accurate predictions to plan entry, descent, and landing (EDL) events~\cite{fonseca2019marswrf}. Electrostatic discharge (ESD) events present an unknown risk to humans and assets on the surface~\cite{ruf2009emission}. Uncertainties in the atmospheric conditions can cause errors in EDL events such as the one that caused the crash of the Schiaparelli Entry Demonstrator Module~\cite{fonseca2019marswrf}. Therefore, a consistent framework of incorporating atmospheric dust opacity observations into global circulation models (GCMs) is required to provide accurate predictions for future EDL events. 

Mars atmospheric models like MarsWRF~\cite{richardson2007planetwrf} and LMD GCM~\cite{forget1999} are used to simulate the impact of dust activity on meteorological quantities like temperature, convective activity, and wind speeds. However, current models cannot resolve the formation of dust storms. Previous studies show that dust distribution is sensitive to Mars climate model parameters and affects the prediction of both thermal and dynamic properties of the atmosphere, including the nature of Hadley circulation and gravity waves~\cite{guzewich2013impact}. Current Mars climate models can produce dust storm events with the aid of prescribed dust profiles from observations of actual dust storm events. The Mars Global Surveyor (MGS) Thermal Emission Spectrometer (TES)~\cite{clancy2010extension,guzewich2013high} and the Mars Reconnaissance Orbiter (MRO) Mars Climate Sounder (MCS)~\cite{heavens2011vertical,heavens2014seasonal} instruments provided limb viewing observations of dust opacity. The dust opacities retrieved from TES and MCS have been used to study weather conditions during dust storms~\cite{guzewich2013impact}. Both MCS and TES have played a key role in providing observations of dust activity over multiple Martian years, including the coverage of global and regional dust storms. TES, Thermal Emission Imaging System (THEMIS)~\cite{smith2019themis}, CRISM~\cite{smith2013vertical}, OMEGA~\cite{d2022vertical}, and MCS have provided observations of dust opacity profiles and optical depth. However, the dust opacity profile retrieval methods of these instruments are distinct from each other, and the use of different dust opacity scenarios produces significant variations in simulation results~\cite{guzewich2013impact,montabone2011reconciling}. 

Radio occultation experiments have been conducted at Mars to study the properties of the ionosphere and neutral atmosphere in the near-surface region~\cite{patzold2016mars,withers2020maven}. In radio science experiments, the spacecraft transmits radio frequency signals to the receiver (Earth or another satellite) through the Martian atmosphere and undergoes bending to atmospheric refraction before reaching the receiver. The phase and Doppler shift measurements at the receiver can be used to retrieve temperature, pressure, and neutral and electron densities of the atmosphere~\cite{fjeldbo1971,hinson1999initial}. Radio science experiments at Mars have provided accurate observations of the neutral atmosphere and ionosphere of Mars. Some important observations are associated with the vertical structure of the ionosphere under various space weather conditions~\cite{mendillo2006effects,peter2021lower}, and investigate the lower atmosphere for trends in gravity wave activity~\cite{creasey2006global} and CO$_{2}$ condensation at the poles~\cite{hu2012mars}. Radio occultation observations can resolve temperatures at the planetary boundary layer (PBL) to understand convective activity and the influence of surface processes and topography on atmospheric dynamics. Past radio occultations have determined the depth of the convective boundary layer and its spatial variation by resolving the temperature structure within the lowest scale height in the atmosphere~\cite{hinson2008depth,fenton2015dust}.

Infrared retrievals have coarse vertical resolutions of about 10--15 km for TES~\cite{clancy2010extension} and 5 km for MCS~\cite{heavens2014seasonal} and CRISM~\cite{smith2013vertical} compared to radio science experiments that provide vertical resolutions up to 1 km above 20 km altitude and $\sim$100 m resolution below 20~km. Infrared (IR) spectrometers are limited by the width of the weighting functions at the wavelengths of the observation. On the other hand, the vertical resolution of radio science is limited by diffraction to the order of the first Fresnel zone diameter. Moreover, algorithms based on small-scale diffraction theory~\cite{karayel1997sub} and the radio holographic method~\cite{jensen2003full} overcome the limitation of vertical resolution by diffraction in radio occultations and provide a significant improvement in measurements of temperature and dust profiles in the planetary boundary layer. 

The effect of dust in climate models could also be implemented using data assimilation~\cite{navarro2014detection}. The lower density of Mars' atmosphere and the lack of oceans pose significant challenges for data assimilation. The radiative time scale of Mars is much shorter than that of Earth and Mars is affected more strongly by the presence of dust aerosols. The assimilation of dust, temperature, and water would improve predictions of synoptic-scale events with GCMs. However, the coarse resolution of currently available temperature and dust opacity data does not provide enough information about the planetary boundary layer. To accurately simulate dust activity, high-resolution data on dust opacity and temperatures are required both in terms of vertical resolution~\cite{montabone2014mars} and spatial coverage. Mars exhibits global-scale structures in dust storms which can be resolved by improving spatial coverage and monitoring them for long durations~\cite{navarro2017challenge}. Currently, radio occultation retrievals of atmospheric temperature profiles provide the highest resolution data required for data assimilation purposes~\cite{montabone2011reconciling,greybush2019ensemble}. 

Amplitude data from the radio occultation of Venus has been used to retrieve the concentration of cloud-forming vapors like SO$_{2}$, and H$_{2}$SO$_{4}$~\cite{jenkins1991results,oschlisniok2021sulfuric}. Radio science experiments on Venus have been used to retrieve SO$_{2}$ and H$_{2}$SO$_{4}$ vapor concentrations, which help us understand the role of meridional circulation in cloud transport~\cite{oschlisniok2021sulfuric}. However, Mars radio occultation experiments from previous missions show that the observations are not sensitive to dust, and any signal loss due to Martian dust has not been reported. The average size of Martian dust particles ranges from \mbox{1 to 2 $\upmu$}m~\cite{lemmon2004,lemmon2019}. Therefore, higher frequency bands with wavelengths of the order of dust size will experience signal loss due to dust activity, which can be recorded in amplitude measurements. It should be noted that spacecraft occultation allows the signal to cover long distances in the atmosphere. Hence, the wavelength of the transmitted signal will need to be larger than the dust size to avoid complete attenuation of the signal, i.e., 30--300 $\upmu$m.

In recent years, the interest in the application of THz technology for Earth and Mars has increased significantly~\cite{chattopadhyay2017terahertz,pradhan2020submillimeter,wedage2022path,wedage2023comparative}. Key applications include satellite telecommunications and remote sensing of trace gases. The THz and far-infrared frequencies are characterized by absorption features of trace gases, in particular water vapor and oxygen. Therefore, we consider a frequency range between 1 and 10 THz (30--300 $\upmu$m) to understand the contribution of dust and water vapor to atmospheric signal loss and test the feasibility of using these bands for dust and water vapor retrieval. The feasibility of orbiter-to-orbiter (O2O) radio occultation (also called cross-link radio occultation) of Mars has been studied in detail~\cite{sweeney2021enabling}. Radio science experiments provide a higher resolution of vertical profiles compared to infrared spectrometer retrievals. They could retrieve high-resolution and accurate measurements of the near-surface environment. Therefore, retrievals of dust opacity from THz-sounding experiments could provide more precise and accurate measurements over current IR dust opacity measurements. Below, we provide an assessment of expected signal loss during occultation and the feasibility of detection with respect to the current state of spacecraft operations and signal processing. Temperature retrieval in radio occultation can resolved at altitudes of $\sim$40 km. However, the objective of the work is to investigate the dust activity in the lower atmosphere of Mars and resolve processes in the planetary boundary layer. Therefore, we do not require temperature measurements above 40 km.

%%%%%%%%%%%%%%%%%%%%%%%%%%%%%%%%%%%%%%%%%%
\section{Materials and Methods}

\subsection{Retrieval of Temperature Profile}
\begin{figure}[H]
\includegraphics[width=15 cm]{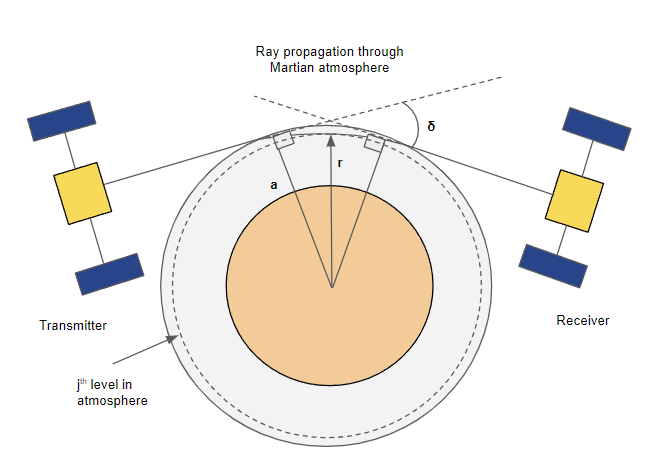}
\caption{Illustration of spacecraft occultation geometry. The ray path from transmitter to receiver undergoes refraction in Martian atmosphere at a bending angle $\delta$ at a given level in atmosphere with a distance from the center of planet $r$ and ray impact parameter $a$.\label{figure1}}
\end{figure}

The temperature profile of the atmosphere can be retrieved from the geometry of spacecraft occultation as described in Figure \ref{figure1}. The geometric parameters are used to calculate the bending angle and refractive index. Furthermore, the refractive index is translated into the number density of the atmosphere using the refractive volume for the Martian CO$_{2}$-N$_{2}$ mixture~\cite{hinson1999initial,fjeldbo1971}.
\begin{eqnarray}
\log(\mu_{j}) & = & \frac{1}{\pi}\int_{a_{j}}^{\infty} \frac{\delta(a)}{\sqrt{a^{2} - a_{j}^{2}}} \,da \\
r_{j} & = & \frac{a_{j}}{\mu_{j}}\\
n_{j} & = & \frac{\mu_{j} -1}{k_{j}},
\end{eqnarray}
where $\mu$ is refractive index, $a$ is the distance of the closest approach, $j$ represents a radio occultation measurement point in the atmosphere, $\delta$ is the bending angle, $k_{j}$ is mean refractive volume that can be expressed as the sum of products of the refractive volume and mixing ratio of constituent gases, $r$ is the radial distance to the planet, and $n$ is the neutral number density of the atmosphere. The temperature and pressure profiles are retrieved from the number density assuming hydro-static balance~\cite{hinson1999initial,patzold2016mars}. 
\begin{eqnarray}
(\frac{\partial P}{\partial z})_{j} & = & -n_{j} m g \\
P(z_{j}) & = & n_{0}k_{B}T_{0} + m\int_{z_{0}}^{z_{j}} n(z')g(z') \,dz' \\
T_{j} & = & \frac{P_{j}}{n_{j}k_{B}}
\end{eqnarray}
where $P$ is pressure, $k_{B}$ is Boltzmann's constant, $T$ is the temperature, $m$ is the gas mean molecular mass, and $g$ is the gravitational acceleration. Thus, $P$ and $T$ can be described as a function of the altitude z. 

\subsection{Atmospheric Losses}
\subsubsection{Gas Absorption}
The main absorbers of the Martian atmosphere in the far-infrared wavelengths are CO$_{2}$, N$_{2}$, H$_{2}$O, CO, and O$_{3}$. In order to assess the contribution of absorbers in the THz band, we compute the maximum expected losses in the range of 0.1 to 10 THz from the HITRAN database~\cite{rothman2021history} for the surface conditions of Mars. Line absorption and broadening for the gases are computed using the Reference Forward Model (RFM)~\cite{dudhia2017reference}. H$_{2}$O and   O$_{2}$ are expected to be the main absorbers below 1 THz~\cite{diao2021comparison}. In addition, the continuum absorption for water vapor is taken into account using the Mlawer--Tobin--Clough--Kneizys--Davies (MT$_CKD$) water vapor continuum model~\cite{mlawer2012development}. We investigate the absorption sources and identify possible sources of background absorption.

\subsubsection{Dust and Water Ice clouds}
Dust particles and water ice aerosols cause absorption and scattering of electromagnetic radiation leading to loss of signal. Expected losses can be estimated from Rayleigh scattering and absorption efficiencies for atmospheric aerosols.
\begin{eqnarray}
Q_{ext} & = & Q_{abs} + Q_{sca} \\
\alpha & = & 4.343Q_{ext}N \pi r_{eff}^{2}
\end{eqnarray}
where $Q_{ext}$ is the extinction efficiency, which can be expressed as the sum of absorption $Q_{abs}$ and scattering efficiencies $Q_{sca}$; $\alpha$ is attenuation coefficient; N is the number density of dust particles; and $r_{eff}$ is the effective dust radius. A projection of expected signal loss during transmission of rays in occultation geometry is given by the following expression~\cite{jenkins1991results}:
\begin{equation}
\tau(r_{j}) = 2\int_{r_{j}}^{\infty} \frac{\alpha(r)}{\sqrt{1 - (a(r_{j})/r\mu(r))^{2}}} \,dr
\end{equation}
where $\tau$ is the expected loss, $\alpha$ is the attenuation coefficient, $a$ is the impact parameter of the ray in the geometry of occultation~\cite{jenkins1991results,oschlisniok2021sulfuric}, and $r$ is the radial distance from the center of the planet to the point of observation. 

\subsection{Retrieval of Dust}
The method for retrieving the dust mass loading utilizes the residual power loss due to the atmosphere similar to the retrieval of sulfuric acid vapor in Venus' atmosphere. However, in the case of Mars, suspended dust particles are the major contributors to signal loss. The residual power loss due to atmospheric attenuation is calculated by subtracting free space loss, refractive defocusing loss, and antenna mispointing errors from the total signal loss~\cite{oschlisniok2021sulfuric}. In the next step, the residual power loss is converted into absorptivity using the inverse Abel transform.
\begin{eqnarray}
T_{tot}(r_{j}) & = & L(r_{j}) + \tau(r_{j}) + m(r_{j}) \\
\alpha(r_{j}) & = & -\frac{n(r_{j})}{\pi \alpha(r_{j})}\frac{d}{da}[\int_{a_{j}}^{\infty} \frac{\tau(a) a}{\sqrt{a^{2} - a(r_{j})^{2}}} \,da] \\
L_{dB} = 10 log_{10}(\frac{I}{I_{0}}) \\
\tau = ln(\frac{I}{I_{0}}) \\
\alpha_{dust}(r_{j}) & = & \alpha(r_{j}) - \alpha_{background}(r_{j}) \\
\alpha_{dust}(r_{j}) & = & 4.343Q_{ext}\rho N \pi r_{eff}^{2} \\
q_{dust} & = & \frac{4}{3} \frac{\rho_{dust} \pi r_{eff}^{3} N}{\rho} 
\end{eqnarray}
where $T_{tot}$ is the total signal loss minus the free space loss, $L$ is refractive loss, $m$ is the mispointing error, and $\tau$ is the atmospheric transmission loss. The absorptivity due to dust ($\alpha_{dust}$) is obtained by removing the contributions from the background gas mixture $\alpha_{background}$. $q_{dust}$ is the mass mixing ratio of dust, which can be obtained from the dust number density N, the average density of the dust particle $\rho_{dust}$, and the density of the atmosphere at that location $\rho$. Decibel losses ($L_{dB}$) can be derived from the signal intensity at the transmitter ($I_{0}$) and receiver (I) ends. The conversion from decibel losses to absorption and scattering losses involves a conversion factor of 4.343, as expressed in Equation (16). The background contributions may be subtracted by assuming an appropriate mixing ratio of absorbers. More details about the concentration of absorbers and their contribution to gas absorption are explained in the next section.

%%%%%%%%%%%%%%%%%%%%%%%%%%%%%%%%%%%%%%%%%%
\section{Results}
\subsection{Instrument Sensitivity and Phase Noise Uncertainty}
A direct assessment of instrument sensitivity to absorption losses in the far-infrared band is not possible. Thereby, we utilize the observed S- and X-band absorptivities in Venus Express radio occultation experiments (VeRas)~\cite{oschlisniok2021sulfuric} as a range to test the potential detection of dust absorptivity by the receiver during occultation mode. Sulfuric acid absorptivities provide the baseline for comparing the magnitude of atmospheric losses that the signal can undergo during transmission through the Martian atmosphere. It provides an optimum range for which the signal undergoes partial attenuation in the atmosphere, enough to be resolved by the receiver.

\begin{figure}[H]
\centering
\includegraphics[width=14 cm]{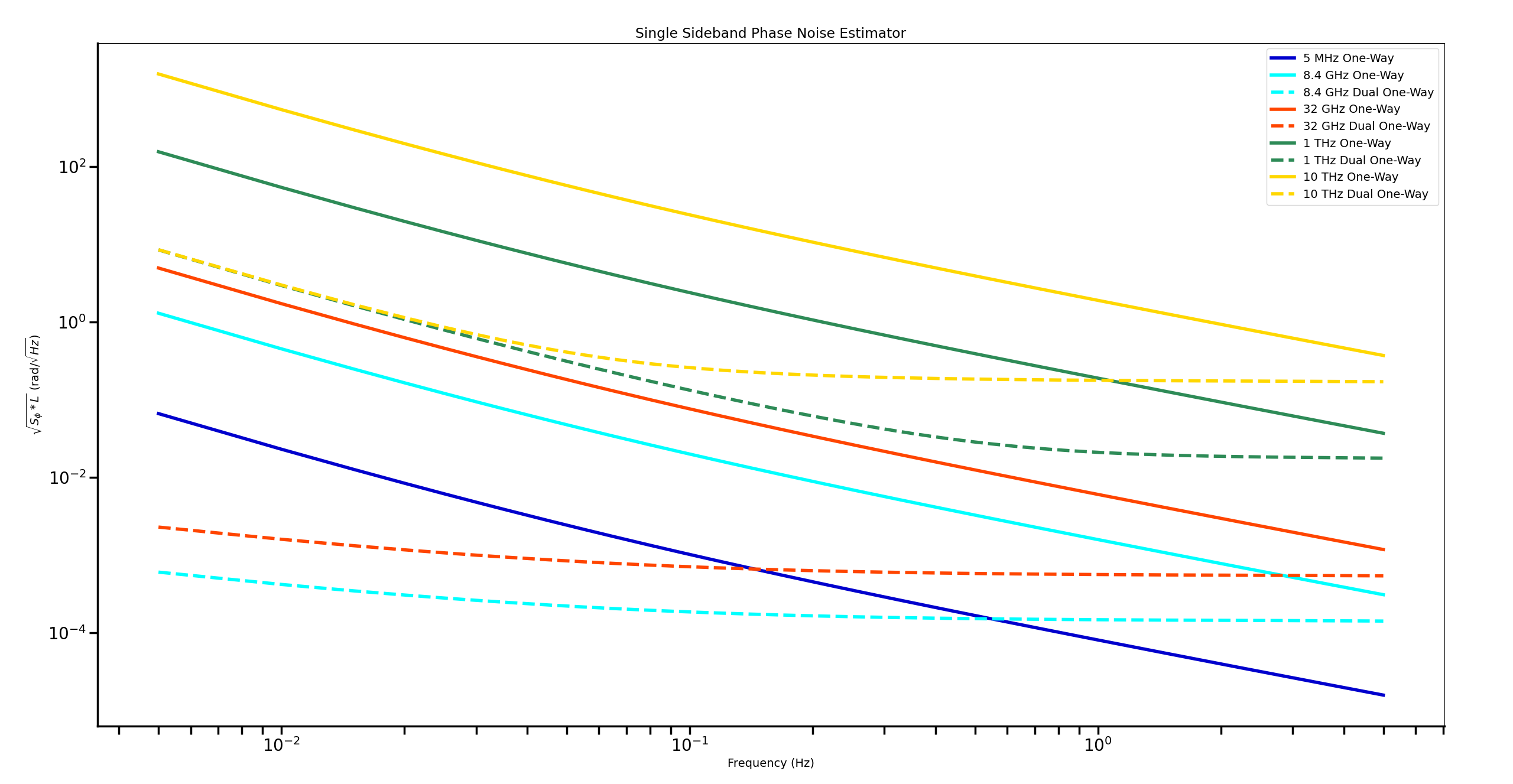}\\
{(\textbf{A})}\\
\includegraphics[width=14 cm]{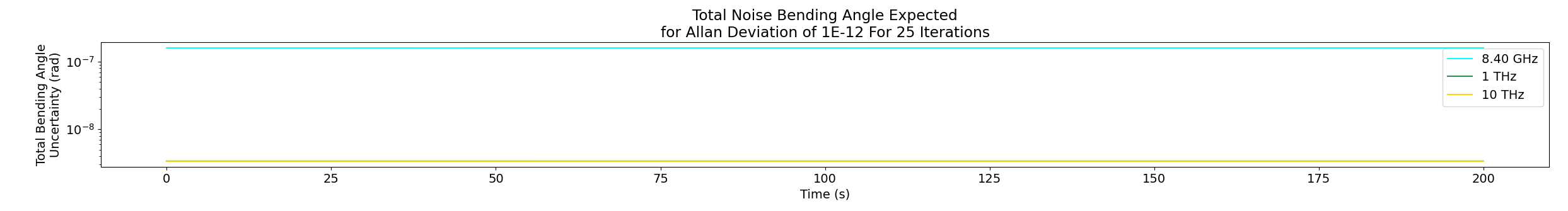}\\
(\textbf{B})\\
\caption{(\textbf{A}) Phase noise estimations for 1 and 10 THz for one-way and dual one-way occultation. The estimations are compared with Iris V2.1 transponder at UHF (5 MHz), X (8.4 GHz) and Ka (32~GHz) band noise; (\textbf{B}) Monte Carlo simulations of expected bending angle noise at 8.4 GHz, 1 THz, and 10 THz at Allan deviation of $10^{-12}$. Expected bending angles for 1 THz and 10 THz do not differ by a significant magnitude. Thereby, the lines overlap with each other.\label{figure2}}
\end{figure}

We do not have any uncertainty quantification for far-infrared occultation instruments. Therefore, we used the sensitivity of the S- and X-band instruments to model the uncertainties in phase noise at the 1 and 10 THz frequencies. In this study, we consider the SmallSat Iris v2.1 transponder proposed for S- and X-band cross-link radio occultation to model uncertainties at THz frequency. As the free space loss is proportional to transmission frequency, higher losses are expected for THz transmission between satellite and Earth. Therefore, we consider the case of O2O cross-link occultation consisting of small satellites as a standard case for THz occultation. Past radio occultation missions relied on a highly accurate USO (Allan deviation of $10^{-13}$ or better), but for a satellite constellation mission, a less accurate clock must be used to realistically fit within the satellite bus. We employ a dual one-way approach (DOW), previously used in gravity range experiments, to sum the phase of two slightly offset frequencies while simultaneously sending and receiving~\cite{thomas1999analysis}. The effect of the noise reduction is shown in Figure \ref{figure2}A, where the radio frequency chosen is shown as a solid line, and the corresponding DOW transmission is shown as a dashed line. For a single clock, the total noise profile for each transmission frequency is drastically lowered with a DOW filter. 

Using the Monte Carlo simulation code developed by~\cite{sweeney2021enabling}, we ran 25 iterations comparing X-band to THz frequency radio occultation observations. The expected noise values combine the thermal uncertainty in the table above along with clock noise generated by an ultra-stable oscillator (USO) power spectral density function as calculated in~\cite{thomas1999analysis}. The chip size clock we plan to use has an Allan deviation of $10^{-12}$. The results of the simulation code (Figure \ref{figure2}B) show that the combined bending angle noise is about two orders of magnitude lower for 1 THz than X band. The lower total noise value allows for higher accuracy when retrieving pressure, temperature and number density values.

\subsection{Atmospheric Absorption}
\begin{figure}[H]
\includegraphics[width=14 cm]{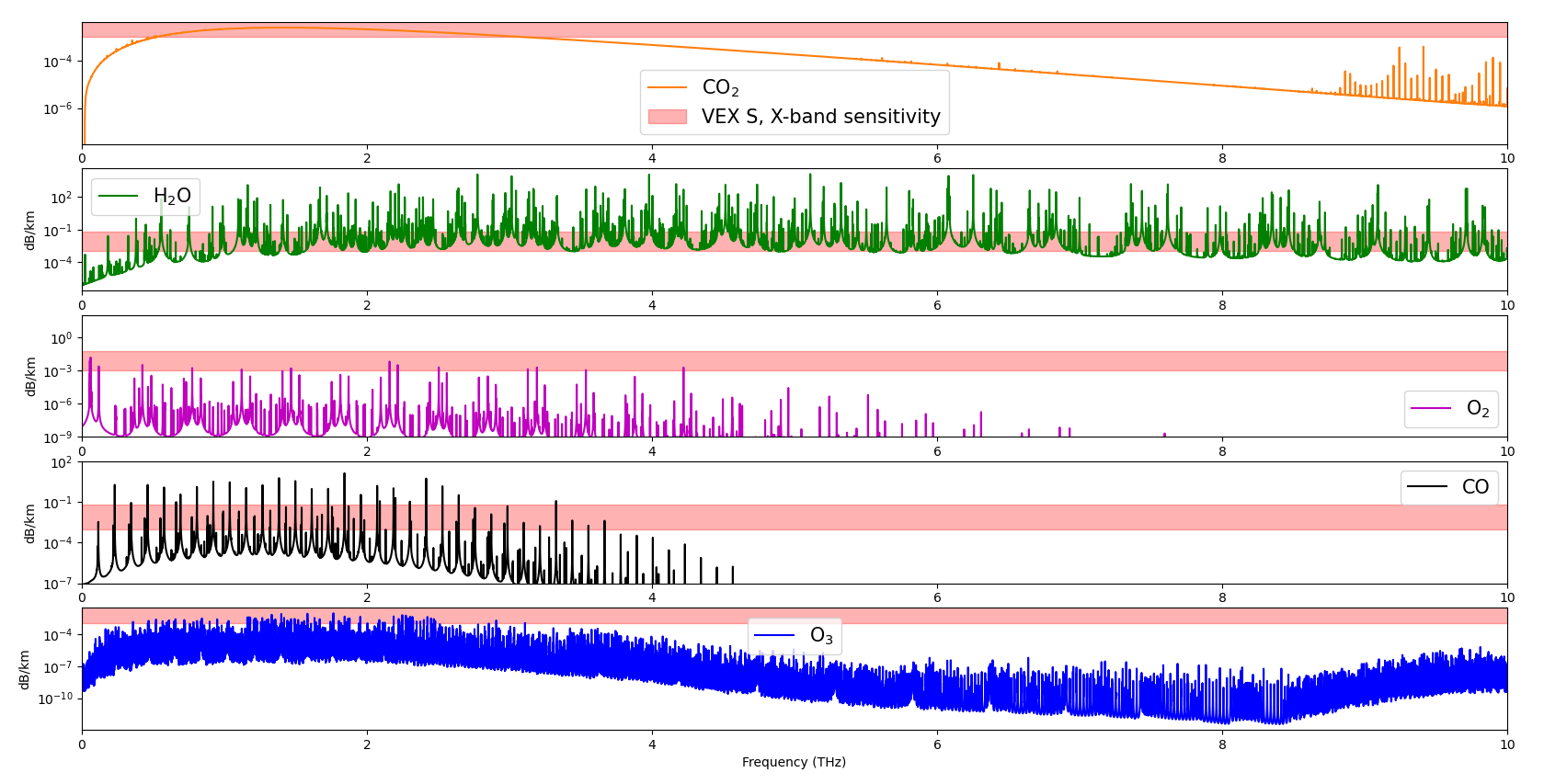}
\caption{Maximum expected signal loss at THz frequencies due to major absorbers in gas phase i.e., CO$_{2}$, CO, H$_{2}$O, O$_{2}$, and O$_{3}$ in Martian atmosphere assuming appropriate mixing ratios~\cite{diao2021comparison} compared with S- and X-band instrument sensitivity. Losses are computed from the absorption coefficient using the HITRAN database.\label{fig1}}
\end{figure}

We provide an estimation of the atmospheric absorption by taking into consideration surface temperature (T = 215 K) and pressure (P = 6.518 mbar) conditions for Mars. These provide the upper bound for gas absorption. The simulated atmospheric losses are computed from the Voigt line shape function using values from the HITRAN database. The pressure-broadening coefficients for each gas correspond to self-broadening. CO$_{2}$-CO$_{2}$ collision-induced absorption~\cite{karman2019update} is also considered along with its line absorption features. Figure \ref{fig1} shows the expected losses due to CO$_2$ (95.32$\%$), H$_2$O (200 ppmv), O$_2$ (0.13$\%$), CO (0.08$\%$), and O$_3$ (0.1 ppmv) assuming mixing ratios of the Martian atmosphere~\cite{diao2021comparison}. H$_{2}$O vapor is expected to show significant absorption losses in the 1--10 THz range. A range of 0.001 to 0.01 dB/km is the optimal absorption to be detected in the occultation experiment~\cite{oschlisniok2021sulfuric}. Moreover, water vapor has a variable concentration in the Martian atmosphere~\cite{fedorova2021multi}. The contribution of water vapor can be resolved with the help of an additional frequency channel in the occultation experiment.

%\vspace{-3pt}

\subsection{Dust and Cloud Losses}
The signal power loss is calculated in four dust storm scenarios~\cite{sheel2016long} using the formulation (Equation (15)) for 
 the dust number density profile~\cite{heavens2011vertical}. In the given equation, $n_{dust}(z)$ refers to the dust density as a function of altitude. The list of parameters for each dust storm scenario is presented in the table below. Here, $n_{s}$ is surface dust density, FH is falloff height, PH is pulse height, FL is falloff length, PT is pulse thickness, and B is the ratio of peak to surface dust~\cite{heavens2011vertical,sheel2016long}. Falloff height refers to the altitude to which dust has been measured, and falloff length is a characteristic length scale where the opacity decays from near-surface dust conditions to zero.
\begin{eqnarray}
n_{dust}(z) & = & n_{s}*A(z) \\
A(z) & = & [1 - e^{- \frac{(z - FH)^{2}}{FL^{2}}}] + B.e^{- \frac{(z - PH)^{2}}{PT^{2}}}
\end{eqnarray}

\begin{figure}[H]
\includegraphics[width=15 cm]{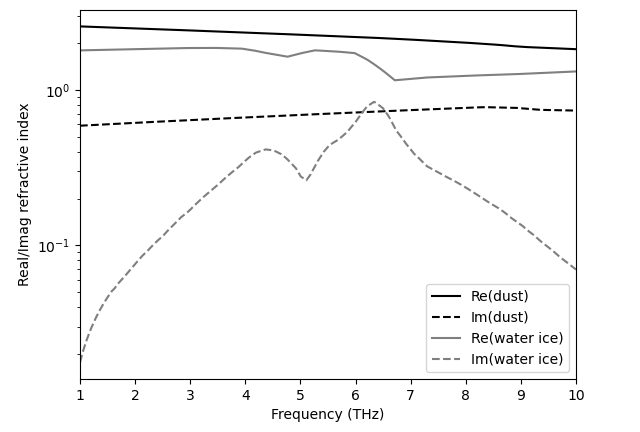}
\caption{Real and imaginary refractive indices for water ice and dust particles at frequencies in the far-infrared ranging from 1 to 10 THz.\label{figure4}}
\end{figure}

We estimate the attenuation due to water ice clouds and compare it with ones from the dust storm scenarios mentioned in Table \ref{tab1}. Signal attenuation is estimated in the Rayleigh scattering regime (Equation (13)) using absorption and scattering coefficients from~\cite{jacobson1999fundamentals}, and the corresponding signal losses at each altitude are computed using the Abel transform for given dust number density profiles (Equation (9)). The complex refractive index of dust between 1 and 10 THz is taken from~\cite{wolff2003constraints}. In the case of water ice, refractive indices are known to be temperature dependent around 10 THz~\cite{warren2008optical}. At 10 THz frequency, i.e., 30 $\upmu$m, the imaginary value of the refractive index changes from 0.03 (T = 176 K) to 0.05 (T = 266 K). The temperature range of the Martian atmosphere shows a wide degree of variation between \mbox{130 to 230 K~\cite{patzold2016mars}}. Therefore, we consider the nominal values for real and imaginary indices~\cite{warren2008optical}. Figure \ref{figure4} shows the magnitude of real and imaginary refractive indices considered for computing the signal loss. 

\begin{table}[H] 
\caption{Parameters for dust activity scenarios. \label{tab1}}
\newcolumntype{C}{>{\centering\arraybackslash}X}
\begin{tabularx}{\textwidth}{m{2.5cm}<{\centering}CCCCCC}
\toprule
\textbf{Dust Activity Scenario}	&\boldmath{$n_{s}$} \textbf{(cm\boldmath{$^{-3}$})}	&\textbf{B}	& \textbf{FH (km)}	& \textbf{FL (km)} & \textbf{PH (km)} & \textbf{PT (km)}\\
\midrule
MY25& 17	& 0.86 & 76 & 12 & 48 & 18\\
MY28    & 12 & 0.86 & 76 & 12 & 48 & 18\\
Regional storm  & 6 & 0.33 & 45 & 9 & 32 & 4\\
No storm & 1.2 & 0.75 & 42 & 12 & 25 & 6\\
\bottomrule
\end{tabularx}
\end{table}
\vspace{-11pt}

In order to draw a comparison between dust and water ice cloud losses with VeRa absorptivity, we calculate the range of expected absorptivity due to water ice clouds and dust storm scenarios. For dust particles, we present the extreme case: (i) global dust storm at MY 28 and (ii) no dust storm based on the dust distribution parameters presented in Table \ref{tab1}. The effective radius of the dust particles is taken to be ~1.5 $\upmu$m~\cite{lemmon2004,lemmon2019} while the radii of the water ice particles are in the range of 1--3 $\upmu$m~\cite{guzewich2019seasonal}. For water ice, the maximum number density is computed from the maximum water ice content expected for the Martian atmosphere, i.e., 1 mg/cm$^{3}$~\cite{kite2021warm}.

\begin{figure}[H]
\begin{adjustwidth}{-\extralength}{0cm}
\centering %% If there is a figure in wide page, please release command \centering
\includegraphics[width=18.5 cm]{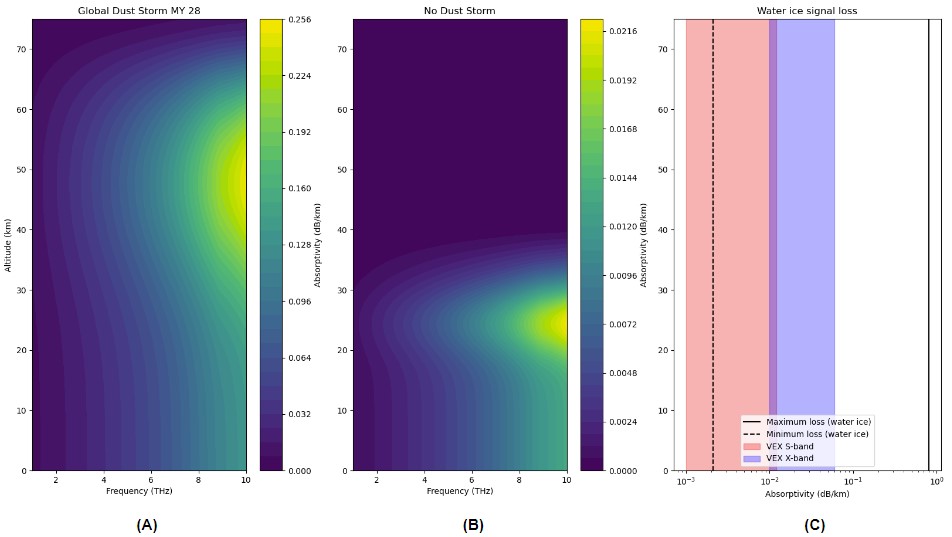}
\end{adjustwidth}
\caption{Dust absorptivity at frequencies from 1 to 10 THz for (\textbf{A}) global dust storm in MY 28 and (\textbf{B}) no dust storm; (\textbf{C}) the range of water ice absorptivities between 1 and 10 THz compared with VeRa S- and X-band absorptivities~\cite{oschlisniok2021sulfuric}.\label{figure5}}
\end{figure}

Figure \ref{figure5}A,B show dust absorptivity corresponding to the range of dust storm absorptivity can range from $\sim$0.001 dB/km to greater than 0.1 dB/km in the case of global dust storms. It shows a wide range encapsulating the S- and X-band absorptivities. Thus, it shows that the range of 1 to 10 THz frequencies provides enough dust absorptivity to be detected from the atmospheric losses during occultation. Water ice absorptivities are also expected to be in the same range which enables both dust and water ice to be detected from signal amplitude measurements during the occultation.
\vspace{-7pt}

\begin{figure}[H]
\hspace{-12pt}\includegraphics[width=15 cm]{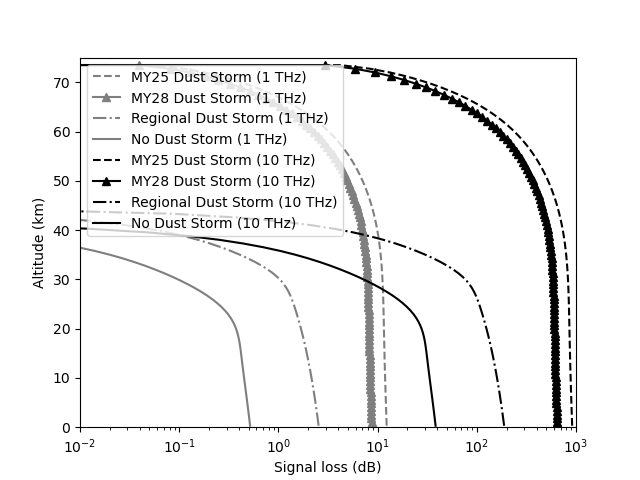}
\caption{Expected signal loss (dB) for Martian dust storms at 1 and 10 THz during signal transmission from occultation estimated from Abel transform of signal attenuation from dust storms.\label{figure6}}
\end{figure}

We also provide an estimation of expected signal loss during a dust storm at 1 THz and 10 THz. The dust distribution for the four scenarios is described in Table \ref{tab1}. We consider the refraction indices derived from a case of an MGS radio science experiment to calculate the signal loss expected over the transmission path. It is to be noted that in a real scenario, the refractive indices and bending angles are expected to vary differently from radio occultation. Figure \ref{figure6} shows that dust storm scenarios can exhibit a wide range of signal loss from $\sim$1 dB to 300 dB. Global dust storms at 10 THz can provide signal losses above 100 dB contributing to signal attenuation. However, at 1 THz, the signal loss due to dust activity is expected to be in the detectable range of $\sim$1--10 dB in the PBL. It shows that the lower end of the frequency band between 1 and 10 THz is suitable for retrieval of dust and water vapor opacities from occultation experiments.

%%%%%%%%%%%%%%%%%%%%%%%%%%%%%%%%%%%%%%%%%%
\section{Discussion and Caveats}

Signal attenuation by global and regional dust storms can be well resolved from atmospheric sounding at far-infrared frequencies. The dust mass mixing ratio and number density can be retrieved from the absorptivity using the inverse Abel transform of atmospheric losses corrected for background absorption (Equations (10)--(16)). The comparison of atmospheric absorption losses with trace species reveals that the background absorbers can contribute to the losses in the range of instrument sensitivity; in particular, CO$_{2}$-CO$_{2}$ CIA, water vapor, and ozone reach magnitudes greater than 0.001 dB/km at frequencies near 1 THz at near-surface conditions. However, in the case of water vapor and ozone, there are windows near 1 THz frequency at which the absorbers do not exhibit significant absorption loss. These bands can be considered as an effective choice of frequency to reduce the contribution of water vapor and ozone below the S- and X-band sensitivity. The contribution of CIA can be removed from the temperature structure and mixing ratio of CO$_{2}$. In addition, multiple frequency channels can be used to deduce the contributions of water vapor and ozone, given that the contributions of absorbers in those bands fall within the instrument sensitivity. Too high loss would lead to complete attenuation of the signal, while too less opacity cannot be resolved from total signal loss. It would allow the receiver to distinguish the effect of water ice and provide vertical distribution of water vapor abundance, given the fact that water vapor is one of the major contributors to gas absorption in the 1--10 THz frequency band (Figure \ref{fig1}). Knowledge of the vertical profile of temperature, dust opacity, and water vapor abundance will provide a better understanding of the evolution of dust storms and cloud formation processes.

The spacecraft occultation method at far-infrared provides higher-resolution data than the infrared limb observations as discussed in the previous section. Cross-link spacecraft occultations are the only option at such high frequencies due to high free space loss and atmospheric absorption loss in Earth's atmosphere. Further, phase measurement and clock noise can be optimized through cross-link occultation. A constellation of small satellites could provide a large set of observations over Martian topography to characterize the development of dust storms. A constellation of four to eight satellites can provide \mbox{$\sim$200--2300 occultation} events per sol depending on the configuration of satellite orbits~\cite{nann2005reconfigurable,sweeney2021enabling}. The vertical resolution is important with regard to understanding the convective and radiative feedback in the near-surface environment and lowest scale heights. Spatial coverage is important to generate large amounts of data required for data assimilation, and cross-link spacecraft occultations can be used to provide the spatial coverage required for data assimilation. In addition, the method will provide dust opacity and water vapor concentrations in the PBL in a wide range of dust opacity conditions. Knowledge of the physical conditions in PBL under varying dust opacity conditions is essential to study the radiative--convective processes and influence global scale models engineering teams use to safely land assets on the surface during EDL. The merit of dust retrieval using far-infrared occultation is that it can provide high-resolution co-located measurements of atmospheric thermal structure and dust opacity, even during high-opacity conditions like regional and planet-encircling dust storms.

In this preliminary study our intent is to demonstrate that the optical properties of Mars atmosphere at terahertz frequency range (1 $\sim$ 10 THz) can be exploited to characterize certain key properties of the planet's atmosphere, without delving into the intricate details of its realistic vertical temperature structure or its spatial and temporal variability.

An extensive discussion of the number of frequency bands required or the necessary bandwidth to effectively
distinguish between atmospheric dust and water is beyond the scope of this preliminary work. However, our approach considers a broad representation, a rough estimate of water ice and aerosol concentrations, which would serve as a first step for an engineering design. The iterative process between
scientific study and engineering advancement is key to making progress. As engineering capabilities expand, so will the depth of
our scientific investigation. 

\textls[-15]{A thorough examination of spectral band selection and spectral inversion techniques---}essential to discerning various confounding
factors--— is the necessary step in the future publication of a comprehensive science study and engineering design. This would entail
considerations of the instrument such as the signal-to-noise ratio (SNR), bandwidth, and its stability as well as establishing a realistic
model of Mars atmosphere, rooted in spatially and temporally varying observations combined with results from global circulation
modeling.

%%%%%%%%%%%%%%%%%%%%%%%%%%%%%%%%%%%%%%%%%%
\section{Conclusions}

Signal attenuation due to dust activity in the 1--10 THz frequency band is investigated under various scenarios of dust activity derived from remote sensing observations. The contribution of dust due to signal attenuation is found to be within the estimates of the observable range from the current state of technology in radio occultation experiments. A method for retrieving the dust mass loading from the signal attenuation in far-infrared transmission that can provide high-resolution data of dust mass mixing ratios in various dust opacity conditions is proposed. This method provides a significant improvement over existing measurements of the opacity of Martian dust. Moreover, the estimation of dust opacity and temperature simultaneously improves the existing capability to predict them during EDL events. The sample return mission~\cite{meyer2022final} will consist of complex operations that would require strong communication support and accurate data on Martian meteorology. An orbiter-based infrastructure for supporting communication can be utilized to retrieve information on the abundance of water vapor, dust opacities, and thermal structure of the surface environment. The high-resolution data retrieved from such a constellation can be leveraged for temperature and dust opacity assimilation in Mars GCMs.

%%%%%%%%%%%%%%%%%%%%%%%%%%%%%%%%%%%%%%%%%%
\vspace{6pt} 

%%%%%%%%%%%%%%%%%%%%%%%%%%%%%%%%%%%%%%%%%%
%% optional
%\supplementary{The following supporting information can be downloaded at:  \linksupplementary{s1}, Figure S1: title; Table S1: title; Video S1: title.}


%%%%%%%%%%%%%%%%%%%%%%%%%%%%%%%%%%%%%%%%%%
\authorcontributions{Conceptualization, A.B. and C.L.; methodology, A.B.; resources, C.L. and D.S.; writing---original draft preparation, A.B.; writing---review and editing, C.L., N.O.R., S.K.A., and D.S. All authors have read and agreed to the published version of the manuscript.}

\funding{This research received no external funding }

\dataavailability{Not applicable.} 

\conflictsofinterest{There is no conflicts of interest.} 

%%%%%%%%%%%%%%%%%%%%%%%%%%%%%%%%%%%%%%%%%%

%% Only for journal Encyclopedia
%\entrylink{The Link to this entry published on the encyclopedia platform.}

\abbreviations{Abbreviations}{
The following abbreviations are used in this manuscript:\\
\noindent 
\begin{tabular}{@{}ll}
MGS & Mars Global Surveyor\\
MRO & Mars Reconnaissance Orbiter\\
TES & Thermal Emission Spectrometer\\
THEMIS & Thermal Emission Imaging System\\
MCS & Mars Climate Sounder\\
CRISM & Compact Reconnaissance Imaging Spectrometer for Mars\\
OMEGA & Observatoire pour la Mineralogie, l’Eau, les Glaces et l’Activité\\
DSN &  Deep Space Network\\
TRL & Technology Readiness Level\\
HITRAN & High-resolution transmission and molecular absorption database\\
PBL & Planetary Boundary Layer \\
DOW & Dual One-Way approach \\
\end{tabular}
}

%%%%%%%%%%%%%%%%%%%%%%%%%%%%%%%%%%%%%%%%%%

%%%%%%%%%%%%%%%%%%%%%%%%%%%%%%%%%%%%%%%%%%
\begin{adjustwidth}{-\extralength}{0cm}

\reftitle{References}
%\bibliographystyle{plain}
%\bibliography{Mars}

\PublishersNote

%%%%%%%%%%%%%%%%%%%%%%%%%%%%%%%%%%%%%%%%%%
\end{adjustwidth}
\end{document}